# Synergistic interplays between the selective electron-phonon coupling, antiferromagnetic fluctuations and charge density wave in the YBa$_2$Cu$_3$O$_x$ cuprate superconductor

[1,2]Chi Ho Wong* and [1]Rolf Lortz*

[1]Department of Physics, The Hong Kong University of Science and Technology, Hong Kong

[2]Department of Industrial and Systems Engineering, The Hong Kong Polytechnic University, Hong Kong

Email(s): chkhwong@ust.hk; lortz@ust.hk

**Abstract:**

This research aims to investigate the synergistic effect between charge density wave, selective electron-phonon coupling under antiferromagnetic fluctuations, as well as the unusual electron distribution observed in ARPES data in YBa$_2$Cu$_3$O$_x$ superconductors (YBCO). By considering their synergistic impact, our model can calculate the superconducting transition temperature T$_c$ of YBa$_2$Cu$_3$O$_x$ as a function of pressure for x = 6.5 and 7 at a reasonable level. Moreover, we have identified a specific antiferromagnetic phonon that contributes significantly to the high T$_c$ observed in YBCO. This finding highlights the significance of these effects in achieving high T$_c$ values. Our study not only identifies an imbalanced charge-density wave effect for triggering selective electron-phonon coupling but also explains why the charge density wave usually occurs around the magnetic copper atoms. Furthermore, our research reveals limitations in the conventional mean-field ab-initio approximation used for antiferromagnetic fluctuations in YBCO. It shows that the dynamic behavior of electrons in YBCO might not be accurately captured by this approximation, as non-uniform magnetic fields under antiferromagnetic fluctuations induce an additional electric potential on electrons across the boundary between non-magnetic to magnetic sites instantaneously. This instantaneous electric potential, in turn, suggest that the influence of the antiferromagnetic phonon-based pairing mechanism might not have been optimized in theory.

**Introduction:**

Cuprate superconductors, also known as high-temperature superconductors (HTS), are a class of materials that exhibit superconductivity at relatively high temperatures [1,2] compared to conventional superconductors [3]. They are composed primarily of copper oxide layers, for instance, YBCO (Yttrium Barium Copper Oxide) superconductors, have been the subject of extensive research and exploration since their discovery in 1986 [1,4]. Despite extensive studies on their crystal structure, phase diagrams and electronic properties, the pairing mechanisms responsible for their high-temperature superconductivity continue to be an open question [5]. In addition, the existence of a pseudogap phase (or the partial energy gap) [6] in these superconductors above the T$_c$ also remains a topic of debate.

Although electron-phonon coupling alone [3] should not be the primary mechanism for unconventional superconductivity, Y.Yagil et al applied Point Contact Spectroscopy PCS measurements to reveal that all of the boson energies observed in YBa$_2$Cu$_3$O$_7$ correspond to specific phonon modes, rather than average phonons [7]. These selective phonons exhibit either softening or hardening below the superconducting transition temperature, where the energy change of the selective phonon is comparable to the superconducting gap [7]. This strongly suggests that phonons should not be ruled out as a part of these bosons [7]. In other words, the electron-phonon coupling in YBCO can be highly selective, meaning that only specific phonon modes are strongly coupled to the electrons [7]. The modulations in the PCS signal are also much stronger than what would be expected from conventional superconductors [7]. These deviations from conventional behavior may be attributed to the presence of magnetic excitations that are strongly coupled to the phonons and are also involved in the pairing mechanism for superconductivity [7].

What kinds of additional excitations are possible to couple with the selective phonon in YBCO? The possible choices could be the following effects. The parent compound of YBCO is an antiferromagnetic (AFM) Mott insulator, with AFM correlations affecting most of the phase diagram under the effect of doping. YBCO also exhibits a charge-density-wave (CDW) order which is characterized by a periodic modulation of the charge density in the crystal lattice [8,9]. Fermiology studies have also revealed the complex nature of its electronic band structure where the Fermi surface always shows d-wave pairing symmetry [10]. Although the electrons on the Fermi surface contribute to the conventional superconductivity only [3], the ARPES technique has revealed that the electrons between the Fermi level $E_F$ and $E_F - E_D$ are missing in unconventional superconductors and surprisingly, these electrons below the Fermi level participate into unconventional superconductivity [11], where this energy range $E_D$ is called the ARPES range. Despite these sophisticated effects have been confirmed to play a role to the unconventional superconductivity, developing a mathematical formulation that captures the interplay between these complex effects and the $T_c$ of YBCO remains a subject of ongoing research. Based on the experimental findings presented Y. Yagil et al [7], we investigate if it is possible to extend their work, in order to explain the high $T_c$ values of YBCO by treating AFM, CDW, d-wave gap anisotropy, and the ARPES range as the additional excitations [8-11].

When AFM fluctuations impose on lattice ions, these AFM fluctuations result in some atoms being influenced by AFM energy, while others are not affected. The atoms under AFM experience slower atomic vibrations, effectively exhibiting a higher effective atomic mass, and eventually the distinct atomic vibration between the AFM and non-AFM lattice sites should emerge a differential out-of-plane phonon [12], which is our trial choice of selective phonon. However, in the context of mean-field approaches such as DFT calculation [13,14], AFM is often treated in a mean field approach, which has a potential risk of masking the effect of the differential out-of-plane phonon. It is important to note that whenever an electron traverses the boundaries between magnetic and non-magnetic sites associated with the differential phonon, it encounters an additional electric potential according to Maxwell's equations [15], while the mean-field ab-initio approximations may not accurately capture the intricacies of this scenario unless a massive calibration work has been done [17].

In 2016, Coh et al. conducted DFT calculations on iron-based superconductors and calibrated a DFT functional to demonstrate the significant amplification effect of electron-phonon coupling

[12] associated with the observed differential out-of-plane phonon in the computation, as later confirmed by experimental results [17]. However, the calibration process for emerging these differential effects in the DFT software can be time-consuming. To simplify this process, an alternative approach called the two-channel model was adopted [16]. This model employs the technique of superposing the impact of the out-of-plane phonon on the upper and lower planes separately before combining their phonon amplitudes. The two-channel model is specifically designed to prevent potential cancellations of charge-density-wave contributions. Notably, our own results [16, 18] are in agreement with the calibrated work by Coh et al., providing additional support for the accuracy of the two-channel model. In 2023, the significance of this effect was demonstrated in successfully simulating the phase diagrams of unconventional superconductors, including $Ba_{1-x}K_xFe_2As_2$, $Ca\ Fe_2As_2$, $SrFe_2As_2$, $LiFeAs$, $NaFeAs$, $FeSe$, and $FeSe/SrTiO_3$ [16, 18].

If the interplay between these intricate effects on electron-phonon coupling has not been thoroughly explored or maximized within the theoretical framework [7], the $T_c$ values of YBCO could have been underestimated. Recognizing this, we develop a theoretical model for calculating the $T_c$ of YBCO as a function of pressure and doping level. Our investigation has the potential to elucidate the key ingredients responsible for the high $T_c$ observed in YBCO. Furthermore, it may potentially facilitate the discovery of new materials exhibiting even higher $T_c$ values.

**Computational Methods:**

We extended the two-channel model [16] to the case of $YBa_2Cu_3O_7$ and $YBa_2Cu_3O_{6.5}$ superconductors with the schematic diagrams in Figure 1. Two possible scenarios are studied in $YBa_2Cu_3O_{6.5}$ in which the oxygen vacancies can be located in Cu-O plane (case A) or Cu-O chain (case B) as shown in Figure 1b and 1c, respectively. Whenever the O atom is above (below) the Cu atom, it refers to the upper (lower) Cu-O channel. In the two-channel model, we define the gain in the electron phonon scattering matrix is $R_{CDW} \sim \dfrac{0.5(DOS_{upper} + DOS_{lower})}{DOS_{both}}$ within the ARPES range [16]. $DOS_{upper}$ represents the mean density of states of electrons for the upper Cu-O channel. $DOS_{lower}$ represent the mean density of states of electrons for the lower Cu-O channel. $DOS_{both}$ is the mean density of states of electrons representing the original crystal structure coexisting upper and lower Cu-O structure. The effective electron-phonon scattering matrix $g$ in the presence of $R_{SDW}$ is $g_{eff} \sim g \cdot R_{CDW}$ [16].

To include all relevant electrons participating in unconventional superconductivity, we set the average electron-phonon scattering matrix $g_{pp'}(E)$ within the ARPES range [11] to be $\left\langle \sum_{E_F - E_{Debye}}^{E_F} \dfrac{g_{pp'}(E')}{\varepsilon'} \right\rangle$. For convenient comparison, we define the ARPES factor as

$$R_{ARPES} \sim \frac{\left\langle \sum_{E_F-E_{Debye}}^{E_F} g_{pp'}(E')/\varepsilon' \right\rangle}{g_{pp'}(E_F)/\varepsilon}.$$ The dielectric constant $\varepsilon'$ is responsible for the screening effect under the complex electrostatic interaction. Apart from these, ellipse equations $p_{angular}(\theta) = \frac{a_{major} b_{minor}}{\sqrt{(b_{minor}^2 - a_{major}^2)\cos^2\theta + a_{major}^2}}$ can be used to mimic the impact of anisotropic superconducting gap under the 4-fold symmetry, where two orthogonally overlapped ellipses (Figure 3b) has occupied an area of $8\int_0^{\pi/4} \frac{1}{2} p_{angular}(\theta)^2 d\theta$ in the momentum space..

In the presence of AFM, the electron-phonon scattering matrix in the spin-unrestricted mode is typically higher than that in the spin-restricted mode by a factor of $R_{AF}$. On the other hand, a comprehensive study in 1977 shows that the effect of AFM on electron-phonon coupling can alternatively be computed by multiplying the exchange enhancement factor to electron phonon coupling in the form of separation of variables [19]. The pressure dependence of AFM interaction can be utilized as the exchange enhancement factor. Under any external or chemical pressure P, the exchange factor can be written as $f(E_{ex}) \sim \frac{[M_{Cu} M_{Cu} E_{co}]_{P>0}}{[M_{Cu} M_{Cu} E_{co}]_{P=0}}$ where the average magnetic moment of the Cu atoms is $M_{Cu}$ and the exchange-correlation energy is $E_{co}$.

Substituting the above effects into the electron-phonon coupling ($\lambda_{PS} = 2\int \alpha_{PS}^2 \frac{F(\omega)}{\omega} d\omega$), in which $F(\omega)$ is the density of states of phonon as a series of frequency $\omega$ [20]. Then the $\alpha_{PS}^2$ becomes $\alpha_{E_F}^2 F(\omega) \cdot R_{AF}^2 \cdot R_{CDW}^2 \cdot R_{APRES}^2 \cdot f(E_{ex})$ where $R_{AF}^2\big|_{P>0} \sim R_{AF}^2\big|_{P=0} f(E_{ex})$. When strong coupling occurs, the renormalized pairing strength $^*\lambda_{PS}$ is substituted in the BCS $T_c$ formula [3]. The ab-initio parameters are listed in the supplementary section.

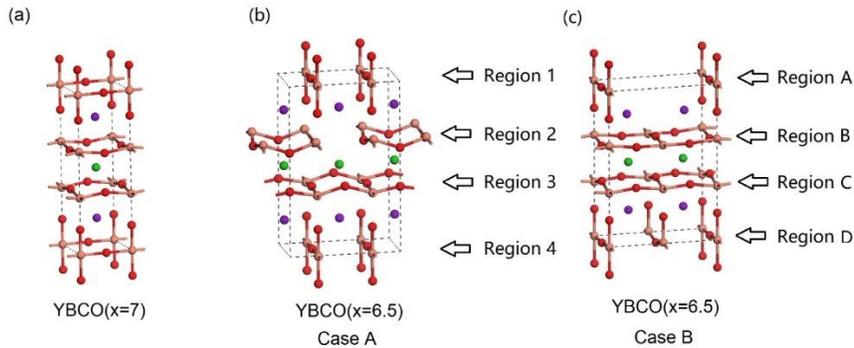

Figure 1: The repeated units of YBCO are drawn after geometric optimization. (a) x = 7; (b) x=6.5 where the oxygen vacancies are located in Cu-O plane; (c) x=6.5 where the oxygen vacancies are introduced into the Cu-O chain.

## Results:

Based on our simulations, the electron phonon coupling is reduced as the pressure rises from 0 to 10GPa. The non-magnetic electron-phonon coupling on the Fermi surface is 0.23 at 0GPa which can be further increased to 0.28 under AFM [21]. Figure 2a illustrates a decrease in the exchange factor of YBCO(x=7) under pressure. The APRES factor displayed in Figure 2b exhibits a drop from 3.8 to 2.8 with increasing pressure from 0 to 10GPa. Furthermore, the $R_{CDW}$ factor experiences an increment from 1.33 to 1.45 under the same pressure conditions in Figure 2c. After considering the d-wave symmetry, we observe that the resultant electron phonon coupling is decreased by ~30% regardless of pressure [21]. After taking an average Debye temperature $T_{Debye}$ in the high and low temperature limits, using the BCS $T_c$ formula is good enough to calculate the $T_c$ of YBCO(x=7) reasonably (see supplementary materials), as plotted in Figure 3a. Figure 3b shows that the theoretical $T_c$ of YBCO(x=7) at 0GPa is only affected by ~15% even when we maximize the effect of gap anisotropy in the entire ARPES range.

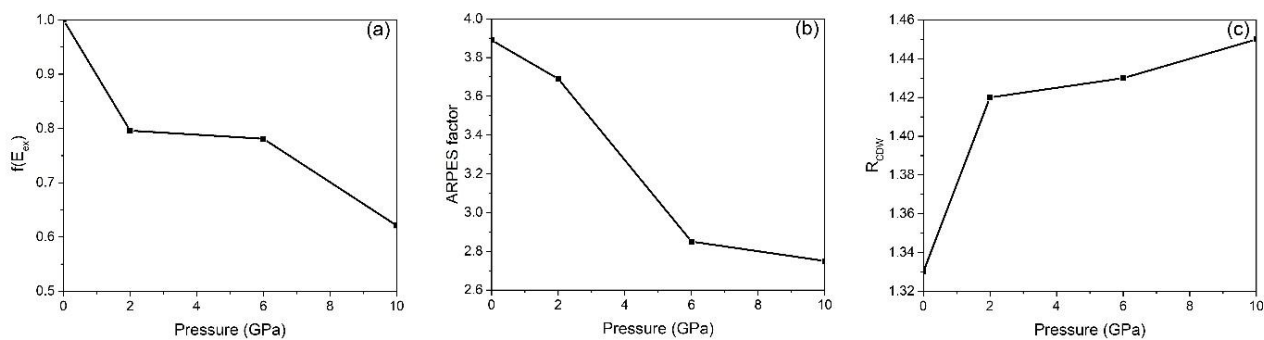

Figure 2: YBCO(x=7) (a) The exchange factor reduces upon compression; (b) The ARPES factor drops under pressure; (c) The effect of charge density wave is stronger when pressure is applied.

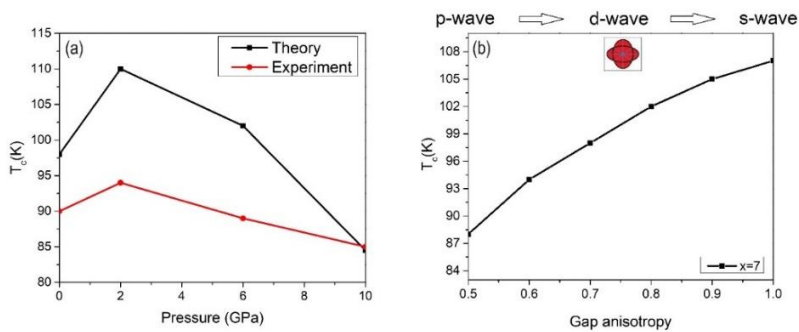

Figure 3: (a) The theoretical and experimental $T_c$ of YBCO(x=7) as a series of pressures [24]. (b) The effect of the anisotropic momentum space on the theoretical $T_c$ of YBCO(x=7, 0GPa). The 4-fold symmetry is represented by the intersection of two red ellipses. To further divide this symmetry into equal partitions, blue lines are used to create 8 regions. Each of these regions has an area of $\int_0^{\pi/4} \frac{1}{2} p_{angular}(\theta)^2 d\theta$. Halving the electron-phonon coupling requires p-wave symmetry [28].

We investigated the $T_c$ of YBCO(x=6.5) at 0GPa in different vacancy locations, where the schematic diagrams are drawn in Figure 1b and 1c. The non-magnetic electron-phonon coupling on the Fermi surface is 0.16 but turning on AFM increases it to 0.21. The d-wave symmetry reduces the AFM-assisted electron-phonon coupling by ~28% regardless of pressure [21]. Interestingly, we observe a significant dependence on the vacancy location in both the $R_{ARPES}$ and $R_{CDW}$ factors. In case A where the vacancy is located in the Cu-O plane (Figure 1b), the two-channel model [16] reveals that the lower channel plays a prominent role in amplifying the electron-phonon scattering matrix. By revising the superconducting electron concentration within the ARPES range, only the lower channel contributes to the amplification effect on the ARPES factor selectively, resulting in almost halving the average $R_{ARPES}$ to 2.12. Distinct amplification of the electron-phonon scattering matrix is also observed in the $R_{CDW}$ factor. Specifically, the average $R_{CDW}$ factor is determined to be 2.02, with the lower channel exhibiting an amplification effect of $R_{CDW}$ ~1.5 times higher than that of the upper channel.

On the other hand, when a vacancy defect is present along the Cu-O chain (Case B, Figure 1c), the upper and lower channels demonstrate a more pronounced distinction in the $R_{ARPES}$ and $R_{CDW}$ factors again. The ARPES factor is determined to be 3.8, indicating a substantial amplification effect on the electron-phonon interaction. If the lattice dynamics is conducted selectively, only the lower channel contributes to a gain in the ARPES factor. The contribution in amplifying the ARPES factor from the lower channel is nearly 7 times higher compared to the upper channel. Meanwhile, the $R_{CDW}$ factor is 1.2, with the lower channel demonstrating the amplification effect ~1.7 times higher than the upper channel. Averaging the results of these two cases, the $<R_{ARPES}>_{AB}$ and $<R_{CDW}>_{AB}$ are 2.96 and 1.61, respectively. Although the $R_{CDW}$ and $R_{ARPES}$ factors vary between Case A and Case B, the corresponding theoretical $T_c$ values are similar. In Case A, the theoretical $T_c$ is determined to be 62K, whereas in Case B, it is calculated to be 67K, both are comparable to the experimental Tc of 59K [24, 25]. The average value of $R_{CDW}$ raises from 1.33 to 1.61 as the value of x changes from 7 to 6.5, but the $R_{APRES}(x=7)$ is higher than $R_{ARPES}(x=6.5)$. In the experiment, when pressure is applied to YBCO with an oxygen content of x=6.5, the experimental $T_c$ increases to 68K [24]. Based on the zero-pressure investigation, there is no significant difference in $T_c$ values between Case A and Case B. Therefore, we focus on studying the compressed YBCO(x=6.5) in Case A alone. The $R_{CDW}$ factor at x=6.5 decreases by ~4% as the pressure goes up to 5 GPa. Using our model, we are able to theoretically estimate a $T_c$ of 79K at 5GPa.

**Discussions:**

There are two ways to tackle a scientific problem: (1) commencing from correlations to investigate the evolution of different forms of order, and (2) identifying existing orders and examining their mutual influences, we have opted for the second approach in this research. While the first method has provided valuable insights into the intertwined nature of the complex orders in YBCO, it has not yet advanced to the subsequent step of offering new insights into understanding the large superconducting gap or $T_c$ to substantiate the impact of individual effects. It should be noted that our intention is not to discredit the first method. Utilizing the first method remains important as it has confirmed the nominated factors triggering unconventional superconductivity. However, our objective is to provide an alternative perspective that explores the influence of different complex orders on the $T_c$ of YBCO. Using the intricately amplified electron-phonon coupling framework as a starting point holds merit. This is supported by the Y. Yagil 's experimental work [7] and another study that observed a substantial electron-phonon coupling with a high interfacial phonon energy (~100meV) in the 100K iron-based superconductor [22]. These findings indicate that the theoretical framework concerning intricately amplified phonon-mediated pairing mechanisms may still require optimization. Furthermore, the confirmation of phonon-mediated mechanism in 200K superconductors under extremely high pressure [26, 27] validates that phonon-mediated mechanism can be used to explain high $T_c$ even above 200K.

In the compressed YBCO, the amplification factors exhibit different trends. In Figure 2, i.e. x = 7, the exchange factor and ARPES factor decrease, while the charge-density-wave factor displays an opposing behavior. The interplay between these factors plays a crucial role in triggering the optimal $T_c$ at 2GPa (Figure 3a). As pressure increases, the phonon-mediated process and spin fluctuations are suppressed, resulting in a reduction of the electron-phonon coupling and exchange factor at high pressures. In addition, more high-energy electrons within the ARPES range are involved under pressure. However, if an excessive number of high-energy electrons is present, they create a stronger screening effect that pales the ARPES factor.

YBCO (x = 6.5) exhibits less chemical pressure compared to YBCO (x = 7), resulting in a decrease in the Debye temperature. This lower Debye temperature leads to a decrease in the ARPES factor when compared to x = 7, as fewer high-energy electrons are observed relatively. The imbalance effect of the charge-density-wave at x = 6.5 can be attributed to the presence of the asymmetric Cu-O-Cu angles. In Figure 1b, the Cu-O-Cu angle along the CuO chain is linear symmetrically (Region 1 or 4), but the Cu-O-Cu angles in the two Cu-O planes (Regions 2 and 3) are not symmetric, triggering distinct $R_{CDW}$ values. This asymmetric out-of-plane phonon is more pronounced in Case B. This is because there is an extra Cu-O-Cu pivot angle in Region D, while no identical pivot angle points upward in Region A. This serves as the first trigger for the imbalanced out-of-plane phonon. The second trigger for this imbalance effect is the difference in the Cu-O-Cu angles between the upper and lower CuO planes in Regions B and C, which can be clearly observed in Figure 1c. Since YBCO (x = 6.5) holds less internal pressure than YBCO (x = 7), the emergence of the out-of-plane phonon for x = 6.5 is more effective. As a result, the $<R_{CDW}>$ increases from 1.33 to 1.61 when x changes from 7 to 6.5. On the other hand, $R_{CDW}$ (x=6.5) decreases by around 4% from 0 GPa to 5 GPa because pressure usually suppresses the formation of the out-of-plane phonon. However, an opposite behavior of $R_{CDW}$ (x=7) is observed under pressure. When YBCO(x=7) is subjected to a pressure of 10 GPa, the average size of the unit cell in a material is reduced from 11.6Å to 11.2Å along the c-axis. However, despite this overall reduction, compression causes an increase in the local distance between the CuO chains and CuO planes by ~0.09% along the c-axis but meanwhile, the separation between adjacent CuO planes becomes closer. This particular increase in local distance along the c-axis enhances the orthogonal lattice vibrations in the CuO chains, contributing to the opposite behaviors.

The impact of the gap anisotropy on the theoretical $T_c$ for both x = 7 and x = 6.5 is found to be relatively small, with an approximate 15% change as shown in Figure 3c. This observation is based on an assumption that the anisotropic momentum space exists throughout the entire ARPES range. Notably, the high-energy electrons distorted by the gap anisotropy should be ineffective. Therefore, the calculated $T_c$ are minimally affected if we focus solely on the gap anisotropy on the Fermi surface [16].

Our findings indicate that considering the selective out-of-plane phonon is enough to explain the high $T_c$ value of YBCO [7]. The conventional ab-initio approximation, which treats AFM fluctuations as a mean AFM field, may not accurately describe the dynamic behavior of electrons [12,17] in YBCO. This is because under AFM fluctuations, the magnetic field is not uniform in space. As electrons move across the lattice points, they experience a time-dependent magnetic field. According to Maxwell's equations, this time-dependent magnetic field can be rewritten as a curl operator of the electric field, which can be further reformulated as an electric potential [15]. Consequently, whenever electrons move across the non-uniform magnetic field, they experience an induced xy potential to boost the electron-ion interaction. In contrast, when a mean-field AFM approximation is employed [13,14], the moving electrons do not encounter an induced xy potential since the time derivative of the magnetic field is zero. Hence, the AFM phonon-mediated mechanism in YBCO may not have been fully optimized when employing conventional spin-unrestricted ab-initio calculations alone.

The choice of using the Debye temperature in the low or high temperature limit depends on the $T_c$ of the system being studied. When $T_c$ is close to 0K, it is appropriate to use the Debye temperature at the low temperature limit. Conversely, when $T_c$ is around 200K [26], the Debye temperature at the high temperature limit should be used. However, in the specific work being discussed, the $T_c$ falls within the range of ~50K to ~100K. Hence, the average Debye temperature is utilized in the BCS $T_c$ formula [3].

Unlike cuprate superconductors, most iron-based superconductors exhibit a spin-density-wave phenomenon characterized by alternating regions of AFM maxima and minima [12, 23]. Applying the two-channel model [16] to the iron-based systems leads to a significant increase in the local magnetic moment of Fe atoms. In contrast, when the two-channel model is applied to YBCO, there is no noticeable change in the magnetic moment of the Cu atoms. This suggests that the two-channel model [16] should be effective in checking whether the sample carries a spin-density-wave or not. Using a first-order AFM fluctuation is found to be sufficient for determining a reasonable trend in the $T_c$ of YBCO under pressure. This is attributed to the absence of a spin-density-wave phenomenon in YBCO which would otherwise further amplify the local AFM maxima through constructive interference-like characteristics, where a higher-order AFM fluctuation may be needed [16]. On the other hand, the $R_{CDW}$ is proportional to the increase of the electronic DOS. This enhanced DOS across the boundaries between non-magnetic and magnetic regions, in turn, can be one of the reasons of forming the CDW phenomenon. However, it poses a huge challenge [12,17] in pinpointing the specific location of the unusual phonons experimentally, where the differential out-of-plane phonon can change its location over time rapidly due to the presence of AFM fluctuations.

Based on our computational model, we have made the following interpretations

- The specific antiferromagnetic phonon, which refers to the instantaneous electron-phonon coupling occurring in specific phonon channels at the boundary between non-magnetic and magnetic lattice sites, should play a role in explaining the high $T_c$ values observed in YBCO.
- When an electron crosses the boundary between non-magnetic and magnetic lattice sites under AFM fluctuations, at this instantaneous moment, the electrons are subjected to an induced electric potential associated with a substantial increase in the electronic DOS at the boundary. This enhanced DOS amplifies the interaction between the electrons and the lattice.
- A large increase in the electronic DOS at the boundary triggers the formation of the CDW phenomenon nearby the magnetic Cu atoms.
- The occurrence of the specific antiferromagnetic phonon is influenced by the positioning of oxygen vacancies within YBCO.
- Most conventional DFT approach may not accurately estimate the instantaneous effect of selective electron-phonon coupling.

Further exploration and investigation are necessary to fully understand the pairing mechanism of YBCO. The pairing mechanism of YBCO continues to be an area of ongoing research, and alternative possibilities for accurately calculating its $T_c$ through different mechanisms are not ruled out.

**Conclusions**

Our research has investigated several factors that influence the superconducting transition temperature of $YBa_2Cu_3O_x$ (x=7 and 6.5) superconductors. We have observed that monitoring the instantaneous differential electron-phonon coupling, ARPES factor, gap anisotropy, exchange enhancement factor and charge-density-wave factor are sufficient to produce the large $T_c$ values associated with the ARPES range. Furthermore, we have identified the specific antiferromagnetic phonon responsible for triggering its large $T_c$ values and it is present across both the non-magnetic and magnetic lattice sites. The presence of an imbalanced charge-density wave in YBCO can be explained by our model. Our findings also suggest that the conventional mean-field ab-initio approximation for antiferromagnetic fluctuations may not fully capture the dynamic behavior of electrons in YBCO, where the non-uniform magnetic field experienced by electrons leads to an induced electric potential as a key factor in boosting the electron-ion interaction.

**Data Availability Statement**

Data are sharable under reasonable request.

**Funding**

This research received no external funding.

Supplementary Sections

1) Ab-initio parameters

The electronic and dielectric properties of the samples are computed using the CASTEP software at the GGA-PBE level of approximation [1]. The maximum self-consistent field (SCF) cycle is set to 1000 with a tolerance of ~$2 \times 10^{-6}$ eV/atom. The reciprocal k-space is sampled with an interval of 0.025 (1/Å). Ultrasoft pseudopotentials are employed in the calculations. To obtain the phonon data, the finite displacement method was employed at the LDA level [2]. A supercell defined by a cutoff radius of 5Å was used, and the dispersion was sampled at an interval of 0.04 (1/Å). The exchange factor is calculated using the spin-unrestricted mode. To ensure the consistency and reliability of our calculations, we utilize available lattice parameters and Debye temperatures $T_{Debye}$ from the literature (if available). Otherwise, the Debye temperatures are computed by CASTEP. While the precise calculation of the pseudopotential μ in unconventional superconductors is still an open question, all Coulomb pseudopotentials are set to 0.2 to enable fair comparisons between different systems.

2) Lattice parameters: [3,4]

At the low-temperature limit, the experimental Debye temperature of YBCO(x=7, 0GPa) is 300K [4], while the Debye temperature from the ab initio calculation [3] at T ~ 0K is 250K.

YBCO(x=7)

| P(GPa) | a(Å) | b(Å) | c(Å) | $T_{Debye(LT)}$(K) | $T_{Debye(HT)}$(K) |
|---|---|---|---|---|---|
| 0 | 3.824 | 3.888 | 11.690 | 250 | 370 |
| 2 | 3.813 | 3.880 | 11.597 | 300 | 443 |
| 6 | 3.791 | 3.868 | 11.443 | 400 | 590 |
| 10 | 3.769 | 3.860 | 11.272 | 500 | 738 |

YBCO(x=6.5)

| P(GPa) | a(Å) | b(Å) | c(Å) | $T_{Debye(LT)}$(K) | $T_{Debye(HT)}$(K) |
|---|---|---|---|---|---|
| 0 (Case A) | 7.508 | 3.914 | 11.707 | 202 | 290 |
| 0 (Case B) | 7.638 | 3.883 | 11.669 | 202 | 290 |

3) The $T_c$ calculation of YBCO(x=7, 0GPa)

$$\lambda_{PS} \sim \lambda_{E_F(d-wave)} \cdot \left[ R_{AF}^2 f(E_{ex}) \right] \cdot R_{ARPES}^2 \cdot R_{CDW}^2$$

$$\lambda_{PS} \sim 0.16 \cdot \left[ 1.06^2 \cdot 1 \right] \cdot 3.89^2 \cdot 1.33^2 = 4.81$$

$$^*\lambda_{PS} = \lambda_{PS} / (\lambda_{PS} + 1) = 0.83, \quad ^*\mu = \mu / (\lambda_{PS} + 1) = 0.034,$$

$T_{Debye(LT)} = 250K$, $T_{Debye(HT)} = 370K$

$$T_{c(LT)} = 1.13 \cdot T_{Debye(LT)} \cdot \exp\left(\frac{-1}{{}^*\lambda_{PS} - {}^*\mu}\right) = 79K \quad T_{c(HT)} = 1.13 \cdot T_{Debye(HT)} \cdot \exp\left(\frac{-1}{{}^*\lambda_{PS} - {}^*\mu}\right) = 117K$$

$<T_c> = (T_{c(HT)} + T_{c(LT)})/2 = 98K$

4) The $T_c$ calculation of YBCO(x=7, 2GPa)

$$\lambda_{PS} \sim \lambda_{E_F(d-wave)} \cdot \left[R_{AF}^2 f(E_{ex})\right] f(E_{ex}) \cdot R_{ARPES}^2 \cdot R_{CDW}^2$$

$$\lambda_{PS} \sim 0.15 \cdot \left[1.06^2 \cdot 0.796\right] \cdot 3.69^2 \cdot 1.42^2 = 3.68$$

${}^*\lambda_{PS} = \lambda_{PS}/(\lambda_{PS}+1) = 0.79$, ${}^*\mu = \mu/(\lambda_{PS}+1) = 0.042$,

$T_{Debye(LT)} = 300K$, $T_{Debye(HT)} = 443K$

$$T_{c(LT)} = 1.13 \cdot T_{Debye(LT)} \cdot \exp\left(\frac{-1}{{}^*\lambda_{PS} - {}^*\mu}\right) = 89K \quad T_{c(HT)} = 1.13 \cdot T_{Debye(HT)} \cdot \exp\left(\frac{-1}{{}^*\lambda_{PS} - {}^*\mu}\right) = 131K$$

$<T_c> = (T_{c(HT)} + T_{c(LT)})/2 = 110K$

5) The $T_c$ calculation of YBCO(x=6.5, 0GPa, Case A)

$$\lambda_{PS} \sim \lambda_{E_F(d-wave)} \cdot \left[R_{AF}^2 f(E_{ex})\right] \cdot R_{ARPES}^2 \cdot R_{CDW}^2 = \lambda_{PS} \sim 0.12 \cdot \left[1.14^2 \cdot 1\right] \cdot 2.12^2 \cdot 2.02^2 = 2.86$$

${}^*\lambda_{PS} = \lambda_{PS}/(\lambda_{PS}+1) = 0.74$, ${}^*\mu = \mu/(\lambda_{PS}+1) = 0.05$,

$T_{Debye(LT)} = 202K$, $T_{Debye(HT)} = 290K$

$$T_{c(LT)} = 1.13 \cdot T_{Debye(LT)} \cdot \exp\left(\frac{-1}{{}^*\lambda_{PS} - {}^*\mu}\right) = 51K \quad T_{c(HT)} = 1.13 \cdot T_{Debye(HT)} \cdot \exp\left(\frac{-1}{{}^*\lambda_{PS} - {}^*\mu}\right) = 74K$$

$<T_c> = (T_{c(HT)} + T_{c(LT)})/2 = 62K$

6) The $T_c$ calculation of YBCO(x=6.5, 0GPa, Case B)

$$\lambda_{PS} \sim \lambda_{E_F(d-wave)} \cdot \left[R_{AF}^2 f(E_{ex})\right] \cdot R_{ARPES}^2 \cdot R_{CDW}^2 = \lambda_{PS} \sim 0.12 \cdot \left[1.14^2 \cdot 1\right] \cdot 3.84^2 \cdot 1.21^2 = 3.76$$

${}^*\lambda_{PS} = \lambda_{PS}/(\lambda_{PS}+1) = 0.78$, ${}^*\mu = \mu/(\lambda_{PS}+1) = 0.042$,

$T_{Debye(LT)} = 202K$, $T_{Debye(HT)} = 290K$

$$T_{c(LT)} = 1.13 \cdot T_{Debye(LT)} \cdot \exp\left(\frac{-1}{{}^*\lambda_{PS} - {}^*\mu}\right) = 55K \quad T_{c(HT)} = 1.13 \cdot T_{Debye(HT)} \cdot \exp\left(\frac{-1}{{}^*\lambda_{PS} - {}^*\mu}\right) = 79K$$

$$<T_c> = (T_{c(HT)} + T_{c(LT)})/2 = 67K$$

Remarks

- Checkpoint 1: When emerging the differential out-of-plane phonon, an obvious increase in the electronic density of states per atom is expected.

- Checkpoint 2: Meanwhile, when transitioning from the spin-restricted mode to the spin-unrestricted mode, the reduction in dielectric constant should be observed.

- Checkpoint 3: The average magnetic moment of the copper (Cu) atoms should decrease under pressure, while the exchange-correlation energy is expected to increase slightly.

The authors are usually supportive in reproducing the calculated $T_c$ values. If fail to pass through the checkpoints, please send your technical request to roywch654321@gmail.com.